\documentclass[%
 reprint,
 amsmath,amssymb,
 aps,
]{revtex4-2}

\usepackage{graphicx}
\usepackage{dcolumn}
\usepackage{bm}
\usepackage{hyperref}
\usepackage{slashed}
\usepackage{float}
\usepackage{tikz-feynman}

\tikzfeynmanset{compat=1.1.0} 

\begin{document}

\preprint{APS/123-QED}

\title{Photon polarization tensor in presence of constant and arbitrary electric field}

\author{L. A. Hern\'andez}
\author{Juan D. Mart\'inez-S\'anchez}%
\affiliation{%
Departamento de F\'isica, Universidad Aut\'onoma Metropolitana-Iztapalapa, Avenida San Rafael Atlixco 186, Ciudad de México 09340, Mexico.}%
\author{R. Zamora}
\email[Corresponding author: ]{rrzamora@uc.cl}
\affiliation{Instituto de Ciencias B\'asicas, Universidad Diego Portales, Casilla 298-V, Santiago, Chile.}
\affiliation{Facultad de Ingenier\'ia, Universidad San Sebasti\'an, Bellavista 7, Recoleta, Santiago, Chile.}


\begin{abstract}
 We compute the photon polarization tensor at one-loop order in the presence of a constant and uniform electric field. Our calculation is carried out for arbitrary field strength using the Schwinger proper-time formalism, and we explicitly derive an expression for the polarization tensor without approximations. We also present a complementary derivation within the strong field approximation. Our main contribution lies in expressing the polarization tensor in terms of a physically motivated tensor basis that ensures transversality and thus preserves gauge invariance. This basis, constructed from the preferred direction defined by the external electric field, makes explicit the breaking of Lorentz symmetry. We verify the consistency of our results by recovering the well-known vacuum polarization tensor in the zero-field limit and by demonstrating agreement between the strong field limit of the general expression and the result obtained directly in the strong field approximation. Interestingly, in the latter case, only one tensor structure survives, corresponding to the transverse dynamics with respect to the field direction, which highlights the dominance of perpendicular modes.
\end{abstract}

\maketitle


\section{Introduction\label{sec1}}

In the framework of quantum electrodynamics (QED), the photon polarization tensor plays a fundamental role in describing how electromagnetic fields affect the propagation of photons in a medium or in vacuum~\cite{Burden:1991uh,Karbstein:2011ja,Karbstein:2013ufa}. This object encodes the quantum corrections to the photon propagator due to vacuum fluctuations, and it is especially relevant when photons interact with external fields, such as electric or magnetic backgrounds. The modifications induced by such fields are not only theoretically interesting, but also crucial for interpreting physical processes where extreme conditions are realized. 

One particularly compelling physical scenario where these effects become significant is in relativistic heavy-ion collisions. In the early stages of such collisions, extremely intense electromagnetic fields are generated, reaching magnitudes far exceeding those attainable in laboratory setups~\cite{Skokov:2009qp,Voronyuk:2011jd,Brandenburg:2021lnj,STAR:2023jdd,Taya:2024wrm,Panda:2024ccj}. These fields are produced mainly in non-central collisions, where the spectators are responsible for such strong electromagnetic fields. These fields can reach hadronic energy scales, making the study of quantum field theoretical effects in strong-field backgrounds highly relevant. Moreover, photons and dileptons produced during the evolution of the quark-gluon plasma (QGP) serve as penetrating probes of the medium, since they interact only electromagnetically and escape largely unaltered from the collision zone~\cite{Alam:1996fd,Cassing:1999es,David:2006sr,Ruan:2014kia}. A proper understanding of their production rates, polarization, and spectral properties requires taking into account how the surrounding electromagnetic fields influence the propagation of virtual and real photons. In this context, the polarization tensor in the presence of an electric field becomes a key quantity, particularly when aiming to model the non-equilibrium, anisotropic conditions of the early-time QGP.

Previous studies have focused extensively on the polarization tensor in the presence of magnetic fields, using techniques such as Schwinger’s proper-time formalism and finite-temperature field theory~\cite{Danielsson:1995rh,Alexandre:2000jc,Sadooghi:2008yf,Chao:2016ysx,Wang:2021ebh,Wang:2021eud,Hattori:2022uzp,Hattori:2022wao,Fukushima:2024ete,Ayala:2020wzl}. However, the case of an electric background~\cite{Karbstein:2011ja,Karbstein:2013ufa,Katkov:2017pif,Marmier:2024rwd}, especially one with arbitrary strength and orientation, has received comparatively less attention due to the technical difficulties associated with the lack of a stable vacuum and the emergence of non-perturbative phenomena such as pair production. A detailed computation of the polarization tensor in such settings is therefore both challenging and necessary.

In this work, we compute the photon polarization tensor in the presence of an arbitrary homogeneous electric field within QED. We perform the analysis at one-loop order, making use of the proper-time method and a careful treatment of tensor structures in order to preserve gauge invariance. This work is organized as follows: In Section~\ref{sec2}, we discuss the effects of the electric field in the Lorentz space and in the charged fermion propagator. Then, we explain the tensor basis used which satisfies the transversality of the theory. With that information at hand, in Subsection~\ref{sec2.1}, we proceed to compute the general case of the photon polarization tensor, and in Subsection~\ref{sec2.2}, we obtain the expression for the polarization tensor in the strong field limit. In Section~\ref{sec3}, we discuss the results obtained. 

\section{Polarization tensor \label{sec2}}

\begin{figure}[t]
    \centering
    \includegraphics[scale=0.35]{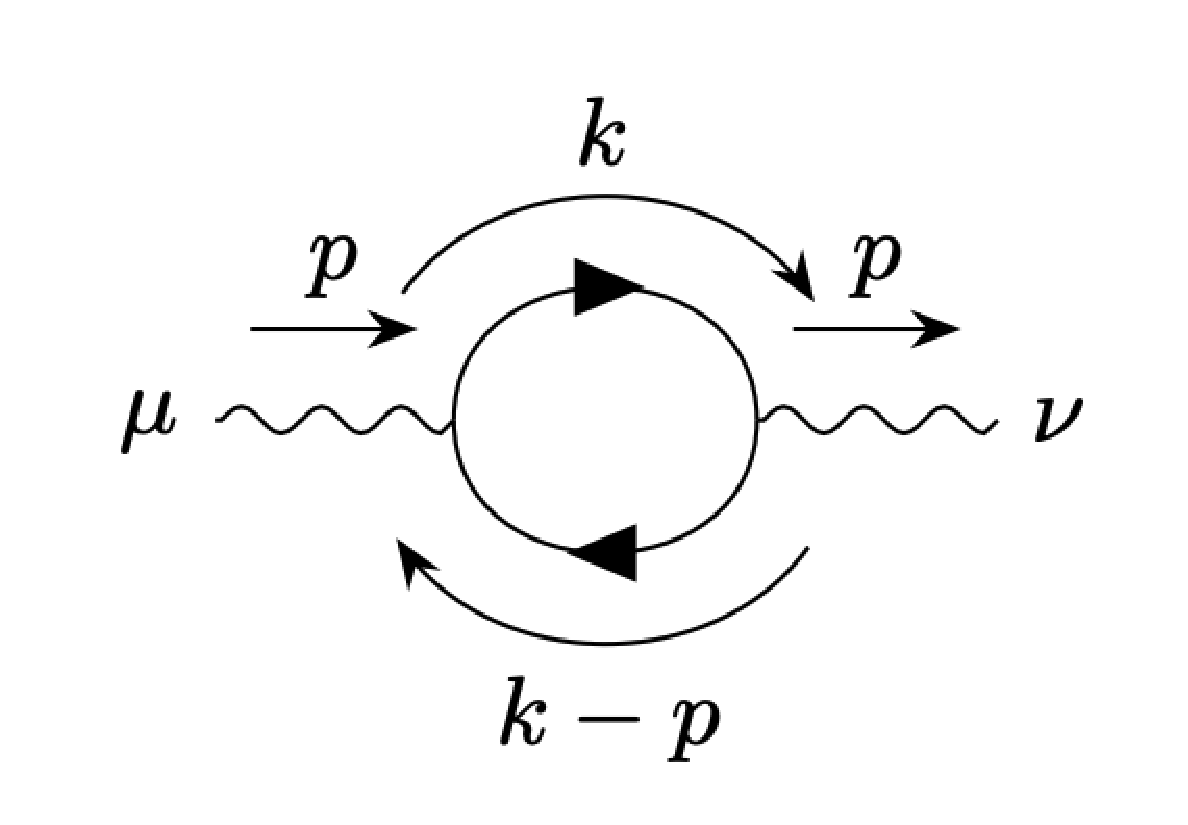}
    \caption{One-loop diagram representing the photon polarization tensor.}
    \label{fig1}
\end{figure}


The photon polarization tensor at one-loop order arises from a virtual fermion–antifermion loop. The corresponding Feynman diagram is shown in  Fig.~\ref{fig1}, and the expression reads
\begin{equation}
    i\Pi^{\mu \nu}(p)=-\frac{1}{2}\int\frac{d^4k}{(2\pi)^4}\text{Tr}[ie\gamma^\mu S^E (k)ie\gamma^\nu S^E(k-p)]+c.c.
    \label{generalpolten}
\end{equation}
Here, $e$ is the electric charge, $\gamma^\mu$ and $\gamma^\nu$ are the Dirac gamma matrices, and $c.c.$ denotes the charge-conjugated contribution. The effect of the external electric field is introduced through the fermion propagator, since the internal lines in the loop correspond to charged particles that couple directly to the background field. The translational invariant contribution of the fermion propagator, written in the Schwinger's proper-time formalism, in the presence of a constant and homogeneous electric field along the direction of $\hat{z}$ is given by
\begin{align}
    S^E(k)&=\int_0^\infty \frac{ds}{\cosh{(|eE|s)}}e^{is(k_\parallel^2\frac{\tanh{(|eE|s)}}{|eE|s}-k_\perp^2-m^2)}\nonumber \\
    &\bigg\{\big( \cosh{(|eE|s)}+s_f\gamma^0\gamma^3\sinh{(|eE|s)}\big)(m-\slashed{k}_\perp)\nonumber\\
    &+\frac{\slashed{k}_\parallel}{\cosh{(|eE|s)}} \bigg\},
    \label{propShcwinger}
\end{align}
where $s_f=\pm 1$ indicates the sign of the electric charge. The momentum decomposition follows the standard separation into longitudinal and transverse components with respect to the electric field direction. Given the Minkowski metric signature $g_{\mu \nu}=\text{diag}(1,-1,-1,-1)$, the conventions adopted throughout this work are
\begin{align}
    g^{\mu\nu}&=g^{\mu\nu}_\parallel+g^{\mu\nu}_\perp, \
 \ \ \ k^2=k_\parallel^2-k_\perp^2, \nonumber \\
    k^2_\perp&=k_1^2+k_2^2, \ \ \ \ \ \ \  k^2_\parallel=k_0^2-k_3^2, \nonumber \\
    \slashed{k}&=\slashed{k}_\parallel-\slashed{k}_\perp, \ \ \ \ \ \ \slashed{k}_\perp=\gamma^1k_1+\gamma^2k_2.
    \label{notation}
\end{align}

Our goal is to compute the photon polarization tensor without imposing any restriction on the strength of the electric field. Nevertheless, the resulting expression must satisfy the transversality condition required by gauge invariance. Since the presence of the external field explicitly breaks Lorentz symmetry, the polarization tensor can be decomposed into three independent transverse tensor structures. Accordingly, we write
\begin{equation}
    \Pi^{\mu\nu}=P_\parallel \Pi_\parallel^{\mu\nu}+P_\perp \Pi_\perp^{\mu\nu}+P_0 \Pi_0^{\mu \nu},
    \label{tensorbasis}
\end{equation}
where the basis tensors are defined as
\begin{align}
    \Pi_\parallel^{\mu \nu}&=g_\parallel^{\mu \nu}-\frac{p_\parallel^\mu p_\parallel^\nu}{p^2_\parallel}, \nonumber \\
    \Pi_\perp^{\mu \nu}&=g_\perp^{\mu \nu}+\frac{p_\perp^\mu p_\perp^\nu}{p^2_\perp}, \nonumber \\
    \Pi_0^{\mu \nu}&=g^{\mu \nu}-\frac{p^\mu p^\nu}{p^2}-\Pi_\parallel^{\mu \nu}-\Pi_\perp^{\mu \nu},
    \label{elementsbasis}
\end{align}
and $P_\parallel$, $P_\perp$ and $P_0$ are the corresponding scalar coefficients. It is important to emphasize that in the present calculation, we do not obtain these coefficients by explicitly projecting the general polarization tensor onto the basis elements defined in Eq.~(\ref{generalpolten}). Instead, we compute the full expression of the polarization tensor from first principles, and only at the end of the calculation express it as a linear combination of the three transverse structures introduced above.

\subsection{General case\label{sec2.1}}

We proceed to calculate the photon polarization tensor for any strength of the electric field; for this purpose we substitute the expression for the propagator in Eq.~(\ref{propShcwinger}) in the general expression for the polarization tensor in Eq.~(\ref{generalpolten}), obtaining 
\begin{align}
    i\Pi^{\mu\nu}(p)&=\frac{e^2}{2}\int \frac{d^4k}{(2\pi)^4}\int_0^\infty \frac{ds e^{is(-k_\perp^2+k_\parallel^2\frac{\tanh(|eE| s)}{|eE| s}-m^2)}}{\cosh(|eE| s)}  \nonumber \\
    &\times \int_0^\infty \frac{ds'e^{is'(-(k-p)_\perp^2+(k-p)_\parallel^2\frac{\tanh(|eE| s')}{|eE| s'}-m^2)}}{\cosh(|eE| s')} \nonumber \\
    &\times \text{Tr}\Bigg[\gamma^\mu\bigg\{ \big( \cosh{(|eE|s)}+s_f\gamma^0\gamma^3\sinh{(|eE|s)}\big)\nonumber \\
    &\times (m-\slashed{k}_\perp)+\frac{\slashed{k}_\parallel}{\cosh{(|eE|s)}} \bigg\}\nonumber \\
    &\times \gamma^\nu \bigg\{ \big( \cosh{(|eE|s')}+s_f\gamma^0\gamma^3\sinh{(|eE|s')}\big)\nonumber \\
    &\times (m-(\slashed{k}-\slashed{p})_\perp)+\frac{(\slashed{k}-\slashed{p})_\parallel}{\cosh{(|eE|s')}} \bigg\}\Bigg].
\end{align}
The first step is to perform the trace over Dirac indices, whose detailed computation is presented in Appendix~\ref{traces}. The resulting expression is
\begin{widetext}
\begin{align}
i\Pi^{\mu \nu}(p)&= \frac{e^2}{2} \int \frac{d^4k}{(2\pi)^4} \int_0^\infty \frac{ds}{\cosh(|eE| s)}  e^{is(-k_\perp^2+k_\parallel^2\frac{\tanh(|eE| s)}{|eE| s}-m^2)} \int_0^\infty \frac{ds'}{\cosh(|eE| s')} e^{is'(-(k-p)_\perp^2+(k-p)_\parallel^2\frac{\tanh(|eE| s')}{|eE| s'}-m^2)} \nonumber \\
&\times \bigg\{\Big(k_\perp^\nu(k-p)_\perp^\mu+k_\perp^\mu(k-p)_\perp^\nu \Big) \mathcal{A}_1 +\Big(k_\perp^\mu(k-p)_\parallel^\nu+k_\perp^\nu(k-p)_\parallel^\mu\Big)\mathcal{A}_2+\Big(k_\parallel^\nu(k-p)_\perp^\mu+k_\parallel^\mu(k-p)_\perp^\nu\Big)\mathcal{A}_3 \nonumber \\
&+\Big(k_\parallel^\nu(k-p)_\parallel^\mu+k_\parallel^\mu(k-p)_\parallel^\nu\Big)\mathcal{A}_4+g^{\mu\nu}\mathcal{A}_5+g^{\mu\nu}_{\parallel}\mathcal{A}_6\bigg\},
\label{poltensoraftertrace}
\end{align}
\end{widetext}
with
\begin{align}
    \mathcal{A}_1&=8\cosh(|eE| s)\cosh(|eE| s')\nonumber \\
    &+8\sinh(|eE| s)\sinh(|eE| s'), \nonumber \\
    \mathcal{A}_2&=-\frac{8\cosh(|eE| s)}{\cosh(|eE| s')}, \nonumber \\
    \mathcal{A}_3&=-\frac{8\cosh(|eE| s')}{\cosh(|eE| s)}, \nonumber \\
    \mathcal{A}_4&=\frac{8}{\cosh(|eE| s)\cosh(|eE| s')}, \nonumber \\
    \mathcal{A}_5&=8m^2\cosh(|eE| s)\cosh(|eE| s')\nonumber \\
&+8m^2\sinh(|eE| s)\sinh(|eE| s') \nonumber \\
&+8\cosh(|eE| s)\cosh(|eE| s')k_\perp\cdot(k-p)_\perp\nonumber \\
&+8\sinh(|eE| s)\sinh(|eE| s')k_\perp\cdot(k-p)_\perp \nonumber \\
&-\frac{8 k_\parallel\cdot(k-p)_\parallel}{\cosh(|eE| s)\cosh(|eE| s')}, \nonumber \\
    \mathcal{A}_6&=-16m^2\sinh(|eE| s)\sinh(|eE| s')\nonumber \\
    &-16\sinh(|eE| s)\sinh(|eE| s')k_\perp\cdot(k-p)_\perp .
\end{align}
The next step is to rewrite the exponentials to obtain Gaussian-like integrals over both the perpendicular and parallel components of the momentum. This requires the following change of variables
\begin{eqnarray}
   && l_{\perp} \equiv k_{\perp}- p_\perp \frac{s'}{s+s} \equiv k_{\perp}- p_\perp \mathcal{B}_1, \nonumber \\
   && l_{\parallel} \equiv k_{\parallel}- p_\parallel \frac{\tanh(|eE| s')}{\tanh(|eE| s)+\tanh(|eE| s')} \nonumber \\
   && \ \ \ \equiv k_{\parallel}- p_\parallel \mathcal{B}_2,  \nonumber \\
\end{eqnarray}
where
\begin{eqnarray}
\mathcal{B}_1 &=& \frac{s'}{s+s'}, \nonumber \\
\mathcal{B}_2&=& \frac{\tanh(|eE| s')}{\tanh(|eE| s)+\tanh(|eE| s')}.
\end{eqnarray}
As a consequence of this change of variables, some straightforward algebraic manipulations are needed in the prefactor multiplying the exponentials. After performing them, we obtain
\begin{widetext}
\begin{eqnarray}
i\Pi^{\mu \nu}(p)= &&4e^2 \int_0^{\infty}ds\int_0^{\infty}ds' 
e^{-ip_{\perp}^2 \mathcal{B}_1s} 
e^{\frac{i p_{\parallel}^2}{|eE|} \mathcal{B}_2 \tanh(|eE|s)}  e^{-i m^2(s+s')} 
\int \frac{d^2l_{\parallel}d^2l_{\perp}}{(2\pi)^4} 
e^{\frac{i}{|eE|}\frac{\tanh{(|eE|s'})}{\mathcal{B}_2}l_{\parallel}^2} 
e^{-i(s+s')l_{\perp}^2}  \nonumber \\
&&\times \Bigg\{ g^{\mu\nu}\Biggl[m^2\mathcal{C}_1+\left(l_{\perp}+p_{\perp}\mathcal{B}_1\right)\left(l_{\perp}+p_{\perp}\mathcal{B}_1-p_\perp\right)\mathcal{C}_1-(l_{\parallel}+p_{\parallel}\mathcal{B}_2)(l_{\parallel}+p_{\parallel}\mathcal{B}_2-p_{\parallel})\mathcal{C}_2\Biggr] \nonumber \\
&&+\left[ \left(l_{\perp}^{\nu}+p_{\perp}^{\nu} \mathcal{B}_1\right)\left(l_{\perp}^{\mu}+p_{\perp}^{\mu} \mathcal{B}_1-p_{\perp}^{\mu}\right)+
\left(l_{\perp}^{\mu}+p_{\perp}^{\mu} \mathcal{B}_1\right)\left(l_{\perp}^{\nu}+p_{\perp}^{\nu} \mathcal{B}_1-p_{\perp}^{\nu}\right)
\right] \mathcal{C}_1 \nonumber \\
&&+ \left[\left(l_{\perp}^{\mu}+p_{\perp}^{\mu}\mathcal{B}_1\right)\left(l_{\parallel}^{\nu}+p_{\parallel}^{\nu}\mathcal{B}_2-p_{\parallel}^{\nu}\right)
+\left(l_{\perp}^{\nu}+p_{\perp}^{\nu}\mathcal{B}_1\right)\left(l_{\parallel}^{\mu}+p_{\parallel}^{\mu}\mathcal{B}_2-p_{\parallel}^{\mu}\right)\right] \mathcal{C}_3 \nonumber \\
&& + \left[\left(l_{\parallel}^{\nu}+p_{\parallel}^{\nu}\mathcal{B}_2\right)\left(l_{\perp}^{\mu}+p_{\perp}^{\mu}\mathcal{B}_1-p_{\perp}^{\mu}\right)
+\left(l_{\parallel}^{\mu}+p_{\parallel}^{\mu}\mathcal{B}_2\right)\left(l_{\perp}^{\nu}+p_{\perp}^{\nu}\mathcal{B}_1-p_{\perp}^{\nu}\right)\right] \mathcal{C}_4 \nonumber \\
&& + \left[\left(l_{\parallel}^{\nu}+p_{\parallel}^{\nu}\mathcal{B}_2\right)\left(l_{\parallel}^{\mu}+p_{\parallel}^{\mu}\mathcal{B}_2-p_{\parallel}^{\mu}\right)
+\left(l_{\parallel}^{\mu}+p_{\parallel}^{\mu}\mathcal{B}_2\right)\left(l_{\parallel}^{\nu}+p_{\parallel}^{\nu}\mathcal{B}_2-p_{\parallel}^{\nu}\right)\right] \mathcal{C}_2\nonumber\\
&&-2 g_{\parallel}^{\mu\nu}(\mathcal{C}_1-1)(m^2+\left(l_{\perp}+p_{\perp}\mathcal{B}_1\right)\left(l_{\perp}+p_{\perp}\mathcal{B}_1-p_\perp\right)) \Bigg\},
\label{Pimunuwithls}
\end{eqnarray}
\end{widetext}
where
\begin{align}
    \mathcal{C}_1&=1 + \tanh(|eE|s)\tanh(|eE|s'), \nonumber \\
    \mathcal{C}_2&=\frac{1}{\cosh(|eE|s)^2\cosh(|eE|s')^2}, \nonumber \\
    \mathcal{C}_3&=-\frac{1}{\cosh(|eE|s')^2}, \ \ \ \ \mathcal{C}_4=-\frac{1}{\cosh(|eE|s)^2}.
\end{align}
We then proceed to perform the integrations over the momentum components. We start with the integration over the perpendicular momentum $l_\perp$, which leads to four master integrals
\begin{eqnarray}
\int d^2l_\perp e^{-i (s+s')l_\perp^2}&=&-i\frac{\pi}{s+s'}, \nonumber \\
\int d^2l_\perp e^{-i (s+s')l_\perp^2}l_\perp^2&=&-\frac{\pi}{(s+s')^2}, \nonumber \\
\int d^2l_\perp e^{-i (s+s')l_\perp^2}l_\perp^\nu l_\perp^\mu&=&\frac{\pi}{2(s+s')^2}g_\perp^{\nu \mu}, \nonumber \\
\int d^2l_\perp e^{-i (s+s')l_\perp^2}l_\perp^\mu l_\perp^\nu&=&\frac{\pi}{2(s+s')^2}g_\perp^{\mu \nu}.
\label{masterintegralsperp}
\end{eqnarray}
Therefore, using the integrals in Eq.~(\ref{masterintegralsperp}), the Eq.~(\ref{Pimunuwithls}) becomes
\begin{widetext}
\begin{eqnarray}
i\Pi^{\mu \nu}(p)&=&4e^2 \int_0^{\infty}ds\int_0^{\infty}ds' 
e^{-ip_{\perp}^2 \mathcal{B}_1s} 
e^{\frac{i p_{\parallel}^2}{|eE|} \mathcal{B}_2 \tanh(|eE|s)}  e^{-i m^2(s+s')} 
\int \frac{d^2l_{\parallel}}{(2\pi)^4} 
e^{\frac{i}{|eE|}\frac{\tanh{(|eE|s'})}{\mathcal{B}_2}l_{\parallel}^2} \frac{\pi}{(s+s')} \nonumber \\
 &\times&\left\{ g^{\mu\nu}\biggl[-im^2\mathcal{C}_1-\frac{1}{(s+s')}\mathcal{C}_1+i p_{\perp}^2 (1-\mathcal{B}_1)\mathcal{B}_1\mathcal{C}_1  \right.  - \mathcal{C}_2 \left(-il_{\parallel}^2 + i p_\parallel^2 \mathcal{B}_2 (1-\mathcal{B}_2) \right)\biggr]  \nonumber \\
&+&\left(\frac{g_\perp^{\mu \nu}}{(s+s')}+2i p_\perp^\mu p_\perp^\nu (1-\mathcal{B}_1)\mathcal{B}_1  \right)
\mathcal{C}_1 + i \mathcal{B}_1 (1-\mathcal{B}_2) (p_\perp^\mu p_\parallel^\nu+p_\perp^\nu p_\parallel^\mu)
\mathcal{C}_3+ i (1-\mathcal{B}_1)\mathcal{B}_2 (p_\parallel^\nu p_\perp^\mu+p_\parallel^\mu p_\perp^\nu)
\mathcal{C}_4 \nonumber \\
&-& \left. i\left(l_\parallel^\nu l_\parallel^\mu+l_\parallel^\mu l_\parallel^\nu-\mathcal{B}_2 (1-\mathcal{B}_2) (p_\parallel^\mu p_\parallel^\nu+p_\parallel^\nu p_\parallel^\mu)\right)
\mathcal{C}_2 - 2g_{\parallel}^{\mu\nu} (\mathcal{C}_1-1)(-im^2-\frac{1}{(s+s')}+ip_\perp^2 (1-\mathcal{B}_1)\mathcal{B}_1 )  \right\}.
\label{Pimunuafterintlperp}
\end{eqnarray}
\end{widetext}
We now carry out the integration over the parallel momentum components $l_\parallel$, which also results in four additional master integrals
\begin{eqnarray}
\int d^2l_\parallel e^{i \frac{\tanh{(|eE|s'})}{|eE|\mathcal{B}_2}l_\parallel^2}&=&\frac{\pi |eE|\mathcal{B}_2}{\tanh{(|eE|s'})} ,\nonumber \\
\int d^2l_\parallel e^{i \frac{\tanh{(|eE|s'})}{|eE|\mathcal{B}_2}l_\parallel^2}l_\parallel^2&=&i\frac{\pi |eE|^2\mathcal{B}_2^2}{\tanh{(|eE|s'})^2}, \nonumber \\
\int d^2l_\parallel e^{i \frac{\tanh{(|eE|s'})}{|eE|\mathcal{B}_2}l_\parallel^2}l_\parallel^\nu l_\parallel^\mu&=&i\frac{\pi |eE|^2\mathcal{B}_2^2}{2\tanh{(|eE|s'})^2}g_\parallel^{\mu \nu} ,\nonumber \\
\int d^2l_\parallel e^{i \frac{\tanh{(|eE|s'})}{|eE|\mathcal{B}_2}l_\parallel^2}l_\parallel^\mu l_\parallel^\nu&=&i\frac{\pi |eE|^2\mathcal{B}_2^2}{2\tanh{(|eE|s'})^2}g_\parallel^{\mu \nu}.
\label{masterintegralspara}
\end{eqnarray}
Substituting the expressions in Eq.~(\ref{Pimunuafterintlperp}), the polarization tensor becomes
\begin{widetext}
\begin{eqnarray}
i\Pi^{\mu \nu}(p)&=&\frac{4e^2|eE|}{(2\pi)^4} \int_0^{\infty}ds\int_0^{\infty}ds' 
e^{-ip_{\perp}^2 \mathcal{B}_1s} 
e^{\frac{i p_{\parallel}^2}{|eE|} \mathcal{B}_2 \tanh(|eE|s)}  e^{-i m^2(s+s')} \frac{\pi^2\mathcal{B}_2}{(s+s')\tanh(|eE|s')} \nonumber \\
&\times&\left\{ g^{\mu\nu}\biggl[-im^2\mathcal{C}_1-\frac{1}{(s+s')}\mathcal{C}_1+i p_{\perp}^2 (1-\mathcal{B}_1)\mathcal{B}_1\mathcal{C}_1  \right.  - \mathcal{C}_2 \left(\frac{|eE|\mathcal{B}_2}{\tanh(|eE|s')} + i p_\parallel^2 \mathcal{B}_2 (1-\mathcal{B}_2) \right)\biggr]  \nonumber \\
&+&\left(\frac{g_\perp^{\mu \nu}}{(s+s')}+2i p_\perp^\mu p_\perp^\nu (1-\mathcal{B}_1)\mathcal{B}_1  \right)
\mathcal{C}_1 + i \mathcal{B}_1 (1-\mathcal{B}_2) (p_\perp^\mu p_\parallel^\nu+p_\perp^\nu p_\parallel^\mu)
\mathcal{C}_3 + i (1-\mathcal{B}_1)\mathcal{B}_2 (p_\parallel^\nu p_\perp^\mu+p_\parallel^\mu p_\perp^\nu)
\mathcal{C}_4 \nonumber \\
&-&\left. i\left(\frac{|eE|\mathcal{B}_2ig_{\parallel}^{\mu \nu}}{\tanh(|eE|s')}-\mathcal{B}_2 (1-\mathcal{B}_2) (p_\parallel^\mu p_\parallel^\nu+p_\parallel^\nu p_\parallel^\mu)\right)
\mathcal{C}_2  + 2g_{\parallel}^{\mu\nu} (\mathcal{C}_1-1)(im^2+\frac{1}{(s+s')}-ip_\perp^2 (1-\mathcal{B}_1)\mathcal{B}_1 )  \right\}.
\label{Pimunuafterintlpara}
\end{eqnarray}
\end{widetext}

Equation~(\ref{Pimunuafterintlpara}) represents the polarization tensor after all internal momentum integrations have been performed, leaving only the integrals over the two Schwinger proper-times. However, before attempting to evaluate these remaining integrals, we reorganize the tensor structures with the goal of expressing the polarization tensor in terms of the basis elements introduced in Eq.~(\ref{tensorbasis}). Therefore, we rewrite the following tensors 
\begin{eqnarray}
 p_\perp^\nu p_\perp^\mu&=&p_\perp^2 \Pi_{\perp}^{\mu \nu} - p_\perp^2 g_{\perp}^{\mu \nu} , \nonumber \\   
  p_\parallel^\nu p_\parallel^\mu&=&-p_\parallel^2 \Pi_{\parallel}^{\mu \nu} + p_\parallel^2 g_{\parallel}^{\mu \nu} ,\nonumber \\
   p_\parallel^\nu p_\perp^\mu+p_\perp^\mu p_\parallel^\nu&=&p^2 \Pi_{0}^{\mu \nu}-p_\perp^2 \Pi_{\parallel}^{\mu \nu} +p_\parallel^2\Pi_{\perp}^{\mu \nu}\nonumber \\ &+&p_\perp^2g_{\parallel}^{\mu \nu}-p_\parallel^2 g_{\perp}^{\mu \nu}, \label{tensores}
\end{eqnarray}
and substitute them into Eq.~(\ref{Pimunuafterintlpara}), getting
\begin{eqnarray}
&&i\Pi^{\mu \nu}(p)=\frac{e^2}{2} \int_0^{\infty}ds \ ds' 
e^{-ip_{\perp}^2 \frac{ss'}{(s+s')}} 
e^{\frac{i p_{\parallel}^2}{|eE|}\frac{\tanh(|eE|s)\tanh(|eE|s')}{\tanh(|eE|s)+\tanh(|eE|s')}} \nonumber \\
&&\times e^{-i m^2(s+s')}  ( P_\parallel \Pi_\parallel^{\mu\nu}+P_\perp \Pi_\perp^{\mu\nu}+P_0 \Pi_0^{\mu \nu} + H_1  g_{\perp}^{\mu\nu}+H_2 g_{\parallel}^{\mu\nu}), \nonumber \\
\label{Pimunufinaltensorstructure}
\end{eqnarray}
where
\begin{widetext}
\begin{eqnarray}
P_{\parallel}&=&i \frac{p_\perp^2|eE|}{4 \pi^2(s+s')^2}\frac{s'\sinh(2 |eE|s)+s\sinh(2 |eE|s)}{\sinh^2(|eE|(s+s'))} - i\frac{p_\parallel^2|eE|}{\pi^2(s+s')}\frac{\sinh(|eE|s)\sinh(|eE|s')}{\sinh^3(|eE|(s+s'))}, \nonumber \\
P_{\perp}&=& i\frac{p_\perp^2 |eE|ss' }{\pi^2 (s+s')^3}\coth(|eE|(s+s'))-i\frac{p_\parallel^2 |eE|}{4\pi^2 (s+s')^2}\frac{s'\sinh(2 |eE|s)+s\sinh(2 |eE|s)}{\sinh^2(|eE|(s+s'))}, \nonumber \\
P_0&=&-i\frac{ |eE|p^2}{4\pi^2 (s+s')^2} \frac{s'\sinh(2 |eE|s)+s\sinh(2 |eE|s)}{\sinh^2(|eE|(s+s'))}, \nonumber \\
H_1&=&-\frac{ |eE|^2}{2\pi^2(s+s')\sinh^2( |eE|(s+s'))}+i \frac{|eE| p_\parallel^2}{4\pi^2(s+s')^2}\frac{s'\sinh(2 |eE|s)+s\sinh(2 |eE|s)}{\sinh^2(|eE|(s+s'))} \nonumber \\
&-&i\frac{ |eE|p_\parallel^2}{2\pi^2(s+s')\sinh^2( |eE|(s+s'))}\frac{ \tanh( |eE|s)\tanh( |eE|s')}{\tanh( |eE|s)+\tanh( |eE|s')} \nonumber \\
&-&i\frac{ |eE|}{2\pi^2(s+s')}\coth( |eE|(s+s'))\Bigl(m^2+\frac{ p_\perp^2 s s'}{(s+s')^2}\Bigr), \nonumber \\
H_2&=& -i\frac{ |eE|}{2\pi^2(s+s')}\frac{\cosh( |eE|(s-s'))}{\sinh(|eE|(s+s'))}\Bigl(m^2-i\frac{1}{s+s'}-\frac{ p_\perp^2 s s'}{(s+s')^2}\Bigr)+i\frac{|eE|p_\parallel^2}{2\pi^2 (s+s')}\frac{\sinh(|eE|s)\sinh(|eE|s')}{\sinh^3(|eE|(s+s'))}\nonumber \\
&-&i \frac{|eE| p_\perp^2}{4\pi^2(s+s')^2}\frac{s'\sinh(2 |eE|s)+s\sinh(2 |eE|s)}{\sinh^2(|eE|(s+s'))}. \label{tensorgeneral}
\end{eqnarray}
\end{widetext}

In Eq.(\ref{Pimunufinaltensorstructure}), five different tensor structures appear, even though only three of them are physically meaningful (as established in Eq.(\ref{tensorbasis})). This apparent inconsistency is resolved by noting that the coefficients $H_1$ and $H_2$ vanish after integration over the two Schwinger proper-times, as demonstrated in Ref.~\cite{ayalajorgito}. Therefore, the final expression for the photon polarization tensor becomes
\begin{widetext}
    \begin{align}
        i\Pi^{\mu \nu}(p)&=\frac{e^2}{2} \int_0^{\infty}ds \ ds'e^{-ip_{\perp}^2 \frac{ss'}{(s+s')}}e^{\frac{i p_{\parallel}^2}{|eE|}\frac{\tanh(|eE|s)\tanh(|eE|s')}{\tanh(|eE|s)+\tanh(|eE|s')}}e^{-i m^2(s+s')} \nonumber \\
      &\times \Bigg \{ \Big( \frac{p_\perp^2|eE|}{4 \pi^2(s+s')^2}\frac{s'\sinh(2 |eE|s)+s\sinh(2 |eE|s)}{\sinh^2(|eE|(s+s'))} - i\frac{p_\parallel^2|eE|}{\pi^2(s+s')}\frac{\sinh(|eE|s)\sinh(|eE|s')}{\sinh^3(|eE|(s+s'))} \Big) \Pi^{\mu\nu}_\parallel \nonumber \\
      &+\Big( i\frac{p_\perp^2 |eE|ss' }{\pi^2 (s+s')^3}\coth(|eE|(s+s'))-i\frac{p_\parallel^2 |eE|}{4\pi^2 (s+s')^2}\frac{s'\sinh(2 |eE|s)+s\sinh(2 |eE|s)}{\sinh^2(|eE|(s+s'))} \Big) \Pi^{\mu\nu}_\perp \nonumber \\
      &+ \Big( -i\frac{ |eE|p^2}{4\pi^2 (s+s')^2} \frac{s'\sinh(2 |eE|s)+s\sinh(2 |eE|s)}{\sinh^2(|eE|(s+s'))} \Big) \Pi^{\mu\nu}_0 \Bigg \}.
      \label{finalPimunufull}
    \end{align}
\end{widetext}
Equation~(\ref{finalPimunufull}) presents the final expression for the photon polarization tensor in the presence of a constant and arbitrary electric field. This result is gauge invariant and, in the limit $|eE| \rightarrow 0$, it agrees with the well-known vacuum result, as shown in Appendix~\ref{apendice2}.

\subsection{Strong field approximation\label{sec2.2}}
We now proceed with a complementary computation of the photon polarization tensor, corresponding to the strong field approximation. In this case, the starting point is given by
\begin{eqnarray}
    i\Pi^{\mu \nu}(p)&=&-\frac{1}{2}\int\frac{d^4k}{(2\pi)^4}\text{Tr}[ie\gamma^\mu \tilde{S}^E (k)ie\gamma^\nu \tilde{S}^E(k-p)]\nonumber\\
    &+&c.c.
    \label{generalpoltenLLL}
\end{eqnarray}
where the propagator for the strong electric field approximation is~\cite{Hattori:2023egw} 
\begin{equation}
    \tilde{S}^E = -2i e^{\frac{ik_{\parallel}^{2}}{|eE|}} \frac{m - \slashed{k}_\perp}{k_{\perp}^{2} + m^{2}}\mathcal{P}_{\pm}, 
    \label{propLLL}
\end{equation}
where $\mathcal{P}_{\pm}$ is the projection operator defined as
\begin{equation}
\mathcal{P}_{\pm} = \frac{1 + s_{f}\gamma^{0}\gamma^{3}}{2},
\label{projectors}
\end{equation}
where $s_f=\pm 1$, as we have defined previously. We proceed to compute the polarization tensor in the strong field approximation. To this end, we begin by substituting Eq.(\ref{propLLL}) into Eq.(\ref{generalpoltenLLL}), obtaining
\begin{eqnarray}
i \Pi^{\mu \nu} (p) &=& -\frac{4e^{2}}{2} \int \frac{d^4k}{(2\pi)^{4}} 
\text{Tr} \Biggl[ \gamma^{\mu} e^{\frac{ik_{\parallel}^{2}}{|eE|}} 
\frac{m - \slashed{k}}{k^{2} + m^{2}} \mathcal{P}_{+} \nonumber \\
&\times&\gamma^{\nu} 
e^{\frac{i(k-p)_{\parallel}^{2}}{|eE|}} 
\frac{m - (\slashed{k} - \slashed{p})_{\perp}}{(k-p)^{2} + m^{2}} \mathcal{P}_{+} \Biggr] \nonumber \\
 &-&\frac{4e^{2}}{2} \int \frac{d^4k}{(2\pi)^{4}} 
\text{Tr} \Biggl[ \gamma^{\mu} e^{\frac{ik_{\parallel}^{2}}{|eE|}} 
\frac{m - \slashed{k}}{k^{2} + m^{2}} \mathcal{P}_{-} \nonumber \\
&\times&\gamma^{\nu} 
e^{\frac{i(k-p)_{\parallel}^{2}}{|eE|}} 
\frac{m - (\slashed{k} - \slashed{p})_{\perp}}{(k-p)^{2} + m^{2}} \mathcal{P}_{-} \Biggr].
\end{eqnarray}
As in the general case, the first step is to perform the trace over Dirac indices. Once this is done, we arrive at the following expression
\begin{eqnarray}
i \Pi^{\mu \nu}(p) &=& \frac{-4 e^{2}}{2} \int \frac{d^{2}k_{\parallel}}{(2 \pi)^{2}} 
e^{\frac{i k_{\parallel}^{2}}{|eE|}} e^{\frac{i (k-p)_{\parallel}^{2}}{|eE|}} 
\nonumber \\
&\times&\int\frac{d^{2}k_{\perp}}{(2\pi)^{2}} 
\frac{1}{(k_{\perp}^{2} + m^{2})((k-p)_{\perp}^{2} + m^{2})} \nonumber \\
&\times& ( 4m^{2} g^{\mu \nu}_{\perp} 
+ 4 ( k_{\perp}^{\mu} (k-p)_{\perp}^{\nu} 
+ g^{\mu \nu}_{\perp} \, k_{\perp}\cdot(k-p)_{\perp} 
\nonumber \\
&+& k^{\nu}_{\perp} (k-p)_{\perp}^{\mu} )).
\end{eqnarray}
Next, we carry out the integration over $k_\parallel$, yielding
\begin{eqnarray}
i \Pi^{\mu \nu}(p) &=& \frac{-e^{2}|eE|}{\pi}
e^{\frac{i p_{\parallel}^{2}}{2|eE|}} I_{\perp}^{\mu\nu}, \label{tensor19}
\end{eqnarray}
with
\begin{eqnarray}
 I_{\perp}^{\mu\nu} &=&\int\frac{d^{2}k_{\perp}}{(2\pi)^{2}} 
\frac{1}{(k_{\perp}^{2} + m^{2})((k-p)_{\perp}^{2} + m^{2})} \nonumber \\
&\times& ( m^{2} g^{\mu \nu}_{\perp} 
+   k_{\perp}^{\mu} (k-p)_{\perp}^{\nu} 
+ g^{\mu \nu}_{\perp} \, k_{\perp}\cdot(k-p)_{\perp} 
\nonumber \\
&+& k^{\nu}_{\perp} (k-p)_{\perp}^{\mu} ).   
\label{20} 
\end{eqnarray}
In order to compute Eq.~(\ref{20}), we use Feynman parametrization. Then, after performing the change of variable $l_{\perp} \equiv k_{\perp} - (1 - x)p_{\perp}$, the resulting expression is
\begin{eqnarray}
   && I_{\perp}^{\mu \nu} = \int_{0}^{1} dx  \int \frac{d^{2} l _{\perp}}{(2\pi)^{2}} \frac{1} {(l_{\perp}^{2} + \Delta)^{2}} 
    \nonumber \\
    &\times& (2 l_{\perp}^{\mu} l_{\perp}^{\nu} + l_{\perp}^{2} g^{\mu \nu}_{\perp} - 2x(1-x)p^ {\mu}p_{\perp}^{\nu}\nonumber \\
    &+&( m^{2} - x(1-x)p_{\perp}^{2}  )g_{\perp}^{\mu \nu}),
    \label{Iperpfunction}\nonumber \\
\end{eqnarray}
where we have defined  $\Delta\equiv x(1-x)p_{\perp}^{2} + m^{2}$. 
To evaluate the integrals over $d^2l_\perp$, we use
\begin{eqnarray}
            \int \frac{d^{d}l}{(2 \pi)^{d}} \frac{1}{(l^{2}  + \Delta)^{n}} &=& \frac{1}{(4 \pi)^{\frac{d}{2}}} \frac{\Gamma(n - \frac{d}{2})}{\Gamma(n)}\left(\frac{1}{\Delta}\right)^{n - \frac{d}{2}},\\
         \int \frac{d^{d}l}{(2 \pi)^{d}} \frac{l^{2}}{(l^{2}  + \Delta)^{n}} &=& \frac{1}{(4 \pi)^{\frac{d}{2}}} \left( \frac{d}{2} \right) \frac{\Gamma(n - \frac{d}{2} -1)}{\Gamma(n)}\nonumber \\
         &\times&\left(\frac{1}{\Delta}\right)^{n - \frac{d}{2} -1},
         \label{l1and2}
    \end{eqnarray}
and the Eq.~(\ref{Iperpfunction}) becomes
\begin{eqnarray}
     I_{\perp}^{\mu \nu} &=& -\frac{1}{4 \pi} g^{\mu \nu}_{\perp} \int_{0}^{1} dx\hspace{0.1cm} \frac{2x (1-x) p_{\perp}^{2}}{\Delta} \nonumber \\
     &-& \frac{1}{4\pi} p_{\perp}^{\mu} p_{\perp}^{\nu} \int_{0}^{1} dx \frac{2x(1-x)}{\Delta} \nonumber \\
      &=& -\frac{1}{4 \pi}{\left( g_{\perp}^{\mu \nu} + \frac{p_{\perp}^{\mu}p_{\perp}^{\nu}}{p_{\perp}^{2}} \right)}\int_{0}^{1} dx \frac{2x (1-x)p_{\perp}^{2}}{\Delta} \nonumber \\
     &=&-\frac{1}{4\pi}\int_{0}^{1} dx \frac{2x (1-x)p_{\perp}^{2}}{ x(1-x)p_{\perp}^{2} + m^{2}} \Pi_\perp^{\mu \nu}.
\end{eqnarray}
Finally, performing the integration over $dx$ yields the following result
\begin{equation}
I_{\perp}^{\mu \nu} = \frac{1}{2\pi} \left(      \frac{4m^{2}\operatorname{arctanh}{\left(  \frac{p_{\perp}}{\sqrt{4m^{2} + p_{\perp}^{2}}} \right)}  }{p_\perp  \sqrt{4m^{2} + p_\perp^{2}}}  - 1  \right)\Pi^{\mu \nu}_{\perp}.
\end{equation}
Therefore, going back to Eq.~(\ref{tensor19}), the photon polarization tensor in the presence of an electric field, within the strong field approximation, is given by
\begin{align}
i\Pi^{\mu \nu}(p)&=\frac{e^2|eE|}{2\pi^2}e^{\frac{ip_\parallel^2}{2|eE|}}\nonumber \\
&\times \left(1-\frac{4m^2 \operatorname{arctanh}\left(\frac{p_\perp}{\sqrt{4m^2+p_\perp^2}}\right)}{p_\perp\sqrt{4m^2+p_\perp^2}} \right)\Pi_\perp^{\mu\nu}.  
\label{Pimunustrongfieldfinal}
\end{align}
It is worth noting that taking the strong electric field limit of Eq.(\ref{finalPimunufull}) leads to Eq.(\ref{Pimunustrongfieldfinal}), thereby reinforcing the consistency of our general result. Further details are provided in Appendix~\ref{apendiceC}.

\section{Conclusions\label{sec3}}

In this work, we have computed the photon polarization tensor at one-loop order in the presence of a constant and uniform electric field along the direction of $\hat{z}$. We carried out both a general computation, without any approximation on the field strength, and a complementary analysis in the strong field approximation. Although the calculation of the photon polarization tensor has been addressed previously in the literature, our contribution lies in the fact that we have explicitly written the tensor in a manifestly transverse form, thereby ensuring gauge invariance. Additionally, our final expression is written in terms of physically motivated tensor structures, constructed using the direction of the external electric field. This representation makes the breaking of Lorentz symmetry evident, as expected in the presence of an external background field. We have checked the consistency of our results by taking two important limits. In the general expression, we verified that taking the electric field to zero recovers the well-known result for the vacuum polarization tensor. Moreover, in the limit of large field strength, the general result coincides with the one obtained in the strong field approximation. An interesting feature of the strong field regime is that only one tensor structure survives, namely the transverse tensor $\Pi^{\mu\nu}_\perp$, which corresponds to the direction perpendicular to the electric field and dominates the dynamics in this limit. These results from this work may prove useful in future theoretical studies aimed at providing a better description of electromagnetic probes in relativistic heavy-ion collision experiments, where strong electric fields can be present during the early stages of the evolution.

\begin{acknowledgments}
Support for this work was received in part by the Consejo Nacional de Humanidades, Ciencia y Tecnología Grant No. CF-2023-G-433. LAH acknowledges support from the DCBI UAM-I PEAPDI 2024, and DAI UAM PIPAIR 2024 projects under Grant No. TR2024-800-00744. R.Z acknowledges support from ANID/CONICYT FONDECYT Regular (Chile) under Grant No. 1241436, and DM-S acknowledges the financial support of a fellowship granted by Consejo Nacional de Humanidades, Ciencia y Tecnología as part of the Sistema Nacional de Posgrados. 
\end{acknowledgments}

\appendix
\section{Traces over Dirac indices \label{traces}}
\begin{widetext}
The first step in the computation of the photon polarization tensor in Eq.~\ref{generalpolten} is to perform the trace over Dirac indices. This can be written as folows
\begin{align}
    &\text{Tr}\Bigg[\gamma^\mu\bigg\{ \big( \cosh{(|eE|s)}+s_f\gamma^0\gamma^3\sinh{(|eE|s)}\big) (m-\slashed{k}_\perp)+\frac{\slashed{k}_\parallel}{\cosh{(|eE|s)}} \bigg\}\nonumber \\
    &\times \gamma^\nu \bigg\{ \big( \cosh{(|eE|s)}+s_f\gamma^0\gamma^3\sinh{(|eE|s)}\big) (m-(\slashed{k}-\slashed{p})_\perp)+\frac{(\slashed{k}-\slashed{p})_\parallel}{\cosh{(|eE|s)}} \bigg\}\Bigg].
\end{align}
We also include the contribution from the conjugate charge and organize the total expression into the following thirteen independent traces
\begin{equation}
T_{1} = m^{2}\cosh{(|eE|s)}\cosh{(|eE|s')}\text{Tr}[\gamma^{\mu}\gamma^{\nu}]=4m^{2}\cosh{(|eE|s)}\cosh{(|eE|s')} g^{\mu \nu},
\label{trace1}
\end{equation}
\begin{equation}
T_{2} = m\cosh{(|eE|s)}\cosh{(|eE|s')}\text{Tr}[\gamma^{\mu}\gamma^{\nu}(\slashed{k} - \slashed{p} )_{\perp}] = 0,
\end{equation}
\begin{equation}
T_{3} = \frac{m\cosh{(|eE|s)}}{\cosh{(|eE|s')}}\text{Tr}[\gamma^{\mu}\gamma^{\nu}(\slashed{k} - \slashed{p} )_{\parallel}] = 0,
\end{equation}
\begin{equation}
T_{4} = m\cosh{(|eE|s)}\cosh{(|eE|s')}\text{Tr}[\gamma^{\mu}\slashed{k}_{\perp}\gamma^{\nu}] = 0,
\end{equation}
\begin{align} 
T_{5} &=\cosh{(|eE|s)}\cosh{(|eE|s')}\text{Tr}[\gamma^{\mu} \slashed{k}_{\perp}\gamma^{\nu}(\slashed{k} - \slashed{p})_{\perp}] \nonumber \\
&= \cosh{(|eE|s)}\cosh{(|eE|s')} k_{\perp}^{\alpha} (k - p)_{\perp}^{\beta} \text{Tr}[\gamma^{\mu} \gamma_{\alpha}^{\perp} \gamma^{\nu} \gamma_{\beta}^{\perp}] \nonumber \\
&= 4 \cosh{(|eE|s)}\cosh{(|eE|s')} [ k_{\perp}^{\mu} (k-p)_{\perp}^{\nu} + g^{\mu \nu}k_{\perp}\cdot (k-p)_{\perp} + k_{\perp}^{\nu}(k-p)_{\perp}^{\mu}],
\end{align}
\begin{align}
T_{6}  &= \frac{\cosh{(|eE|s)}}{\cosh{(|eE|s')}}\text{Tr}[ \gamma^{\mu} \slashed{k}_{\perp} \gamma^{\nu}(\slashed{k} - \slashed{p})_{\parallel}]\nonumber \\
&=\frac{\cosh{(|eE|s)}}{\cosh{(|eE|s')}}k_{\perp}^{\alpha}(k-p)_{\parallel}^{\beta}\text{Tr}[\gamma^{\mu}\gamma_{\alpha}^{\perp}\gamma^{\nu}\gamma_{\beta}^{\parallel}]\nonumber \\ 
&= \frac{4 \cosh{(|eE|s)}}{\cosh{(|eE|s')}} ( k_{\perp}^{\mu} (k-p)_{\parallel}^{\nu} + k_{\perp}^{\nu}(k-p)_{\parallel}^{\mu}), 
\end{align}
\begin{equation}
T_{7} =m^{2}\sinh{(|eE|s)}\sinh{(|eE|s')} \text{Tr}[\gamma^{\mu}\gamma^{0}\gamma^{3}\gamma^{\nu}\gamma^{0}\gamma^{3}]= 4 m^{2} \sinh{(|eE|s)}\sinh{(|eE|s')} ( g^{\mu \nu} - 2g^{\mu \nu}_{\parallel}),
\end{equation}
\begin{equation}
T_{8} = m\sinh{(|eE|s)}\sinh{(|eE|s')}(k-p)_{\alpha}^{\perp} \text{Tr}[\gamma^{\mu}\gamma^{0}\gamma^{3}\gamma^{\nu}\gamma^{0}\gamma^{3}\gamma_{\perp}^{\alpha}] = 0,
\end{equation}
\begin{equation}
T_{9} = m\sinh{(|eE|s)}\sinh{(|eE|s')}k_{\alpha}^{\perp}\text{Tr}[\gamma^{\mu}\gamma^{0}\gamma^{3}\gamma^{\alpha}_{\perp}\gamma^{\nu}\gamma^{0}\gamma^{3}] = 0,
\end{equation}
\begin{align}
T_{10}  & = \sinh{(|eE|s)}\sinh{(|eE|s')}\text{Tr}[\gamma^{\mu}\gamma^{0}\gamma^{3}\slashed{k}_{\perp}\gamma^{\nu}\gamma^{0}\gamma^{3}( \slashed{k} - \slashed{p})_{\perp}]\nonumber\\ &= \sinh{(|eE|s)}\sinh{(|eE|s')} k_{\alpha}^{\perp} (k-p)_{\beta}^{\perp}\text{Tr}[\gamma^{\mu}\gamma^{0}\gamma^{3}\gamma^{\alpha}_{\perp}\gamma^{\nu}\gamma^{0}\gamma^{3}\gamma^{\beta}_{\perp}] \nonumber \\ &=4\sinh{(|eE|s)}\sinh{(|eE|s')} ( -2 k_{\perp} \cdot (k-p)_{\perp} g^{\mu \nu}_{\parallel}+  k_{\perp}^{\nu} (k - p)_{\perp}^{\mu} + k^{\mu}_{\perp}(k-p)_{\perp}^{\nu} 
+ k_{\perp}\cdot (k - p)_{\perp} g^{\mu \nu}),
\end{align}
\begin{equation}
T_{11} = \frac{m \cosh{(|eE|s')}}{\cosh{(|eE|s)}}\text{Tr}[\gamma^{\mu}\slashed{k}_{\parallel}\gamma^{\nu}] = 0,
\end{equation}
\begin{align}
T_{12} &=\frac{\cosh{(|eE|s')}}{ \cosh{(|eE|s)} } \text{Tr}[ \gamma^{\mu} \slashed{k}_{\parallel} \gamma^{\nu} (\slashed{k} - \slashed{p} )_{\perp} ] \nonumber \\ 
&=\frac{\cosh{(|eE|s')}}{ \cosh{(|eE|s)} }k_{\parallel}^{\alpha}(k-p)_{\perp}^{\beta}\text{Tr}[\gamma^{\mu}\gamma_{\alpha}^{\parallel}\gamma^{\nu}\gamma_{\beta}^{\perp}] \nonumber \\
&= \frac{4 \cosh{(|eE|s')}}{\cosh{(|eE|s)}} (k_{\parallel}^{\mu} (k-p)_{\perp}^{\nu} + k_{\parallel}^{\nu}(k-p)_{\perp}^{\mu}),
\end{align}
\begin{align}
T_{13} & = \frac{1}{ \cosh{(|eE|s)}\cosh{(|eE|s')} }\text{Tr}[\gamma^{\mu} \not{k}_{\parallel}\gamma^{\nu} ( \not{k} - \not{p} )_{\parallel} ] \nonumber \\ 
 & =\frac{1}{ \cosh{(|eE|s)}\cosh{(|eE|s')} } k_{\alpha}^{\parallel}(k-p)_{\beta}^{\parallel} \text{Tr}[\gamma^{\mu}\gamma^{\alpha}_{\parallel}\gamma^{\nu}\gamma^{\beta}_{\parallel}]  \nonumber \\ 
&= \frac{4}{\cosh{(|eE|s)}\cosh{(|eE|s')}} (k_{\parallel}^{\mu}(k-p)_{\parallel}^{\nu} 
- g^{\mu \nu}k_{\parallel}\cdot (k-p)_{\parallel} + k_{\parallel}^{\nu}(k-p)_{\parallel}^{\mu}).
\label{trace13}
\end{align}

The results from Eqs.~(\ref{trace1}) through~(\ref{trace13}) allow us to recover the expression for the polarization tensor shown in Eq.~(\ref{poltensoraftertrace}) of the main text, where the derivation proceeds.
\end{widetext}

\section{Vacuum limit $|eE|\rightarrow 0$ \label{apendice2}}

To verify the consistency of our result, we compute the limit in which the electric field strength vanishes. We start from Eq.~(\ref{finalPimunufull}) and take the limit $|eE| \rightarrow 0$ in each term that depends explicitly on the field strength. In the exponent, we have
\begin{equation}
    \lim_{|eE| \to 0} \frac{\tanh(|eE|s)\tanh(|eE|s')}{|eE|(\tanh(|eE|s)+\tanh(|eE|s'))} = \frac{s s'}{ s+s'},
\end{equation}
and the $|eE|$-dependent parts of the coefficients associated with the basis tensor structures are
\begin{eqnarray}
\lim_{|eE| \to 0} |eE| \frac{s'\sinh(2 |eE|s)+s\sinh(2 |eE|s')}{\sinh^2(|eE|(s+s'))} &=& \frac{4s s'}{ (s+s')^2}, \nonumber \\
\lim_{|eE| \to 0} |eE|  \frac{\sinh(|eE|s)\sinh(|eE|s')}{\sinh^2(|eE|(s+s'))} &=& 0, \nonumber \\
\lim_{|eE| \to 0} |eE|  \frac{\sinh(|eE|s)\sinh(|eE|s')}{\sinh^3(|eE|(s+s'))} &=& \frac{s s'}{(s+s')^3}, \nonumber \\
\lim_{|eE| \to 0} |eE|  \coth(|eE|(s+s')) &=& \frac{1}{s+s'}. \nonumber \\
\end{eqnarray}
Substituting the results from the approximations above into Eq.~(\ref{tensorgeneral}), we obtain the coefficients corresponding to each of the tensor structures
\begin{eqnarray}
P_\parallel &=& -i \frac{p^2 s s'}{\pi^2(s+s')^4},   \nonumber \\
P_\perp &=&  -i \frac{p^2 s s'}{\pi^2(s+s')^4},   \nonumber \\
P_0 &=&  -i \frac{p^2 s s'}{\pi^2(s+s')^4}.  
\end{eqnarray}
Finally, the expression for the polarization tensor is
\begin{eqnarray}
&&i\Pi^{\mu \nu}(p)=\frac{e^2}{2} \int_0^{\infty}dsds' 
e^{i p^2 \frac{ss'}{s+s'}}e^{-i m^2(s+s')} 
 \nonumber \\
&&\times \left( -i \frac{p^2 s s'}{\pi^2(s+s')^4}   ( \Pi_\parallel^{\mu\nu}+ \Pi_\perp^{\mu\nu}+ \Pi_0^{\mu \nu} )\right) \nonumber \\
 &=& \frac{e^2}{2} \int_0^{\infty}dsds' 
e^{i p^2 \frac{ss'}{s+s'}}e^{-i m^2(s+s')} 
 \nonumber \\
&&\times \left(-i \frac{p^2 s s'}{\pi^2(s+s')^4} \left(g^{\mu \nu}-\frac{p^\mu p^\nu}{p^2}\right)\right).
\label{PimunueE0}
\end{eqnarray}
The result obtained in Eq.~(\ref{PimunueE0}) is consistent with standard results in the literature, as it recovers the transverse tensor structure in four dimensions, which emerges as a consequence of restoring Lorentz symmetry.

\section{Strong field limit $|eE|\rightarrow \infty$ \label{apendiceC}}
In this appendix, we verify that our general result reproduces the expression obtained in the strong field approximation, which was derived starting from the propagator of charged particles in the strong field limit. To this end, we start from Eq.~(\ref{finalPimunufull}) and take the limit $|eE| \rightarrow \infty$. In the exponential, we find
\begin{equation}
   \lim_{|eE| \to \infty} \frac{\tanh(|eE|s)\tanh(|eE|s')}{|eE|(\tanh(|eE|s)+\tanh(|eE|s'))} = \frac{1}{2 |eE|}, 
   \label{aproxfuerte1}
\end{equation}
and the $|eE|$-dependent terms within the coefficients of the basis tensor structures are
\begin{eqnarray}
\lim_{|eE| \to \infty} |eE| \frac{s'\sinh(2 |eE|s) + s\sinh(2 |eE|s')}{\sinh^2(|eE|(s + s'))} &=& 0, \nonumber \\
\lim_{|eE| \to \infty} |eE| \frac{\sinh(|eE|s)\sinh(|eE|s')}{\sinh^2(|eE|(s + s'))} &=& 0, \nonumber \\
\lim_{|eE| \to \infty} |eE| \frac{\sinh(|eE|s)\sinh(|eE|s')}{\sinh^3(|eE|(s + s'))} &=& 0, \nonumber \\
\lim_{|eE| \to \infty} |eE| \coth(|eE|(s + s')) &=& |eE|. \nonumber \\ 
\label{aproxfuerte2}
\end{eqnarray}
Using the results from Eqs.~(\ref{aproxfuerte1}) and~(\ref{aproxfuerte2}), and substituting them into Eq.~(\ref{tensorgeneral}), we obtain the coefficients corresponding to each of the tensor structures
\begin{eqnarray}
  P_{\parallel}&=& 0, \nonumber \\
  P_{\perp}&=&  i\frac{p_\perp^2 |eE|ss' }{\pi^2 (s+s')^3}, \nonumber \\
    P_0 &=&0.
\end{eqnarray}
Consequently, the polarization tensor takes the form
\begin{eqnarray}
i\Pi^{\mu \nu}(p)&=&\frac{e^2|eE|p_\perp^2}{2\pi^2}\int_0^\infty dsds'e^{-ip_\perp^2 \frac{ss'}{s+s'}}e^{\frac{ip_\parallel^2}{2|eE|}}e^{-im^2(s+s')}\nonumber \\
&\times& \frac{ss'}{(s+s')^3},
\end{eqnarray}
and after performing the change of variables $s = u(1 - v)$ and $s' = uv$, we obtain
\begin{eqnarray}
i\Pi^{\mu \nu}(p)&=& \frac{e^2|eE|p_\perp^2}{2\pi^2} \int_0^{1}du\int_0^{\infty}dv 
e^{i(p_\perp^2 u v (1-v)-m^2u)}e^{\frac{ip_\parallel^2}{2|eE|}} 
 \nonumber \\
&\times& iu\frac{ u^2(1-v) }{ u^3}\Pi_\perp^{\mu\nu}.
\end{eqnarray}
Integrating over the variable $u$, we get
\begin{equation}
i\Pi^{\mu \nu}(p)=-\frac{e^2|eE|}{2\pi^2}e^{\frac{ip_\parallel^2}{2|eE|}} \int_0^1 dv\frac{p_\perp^2 v(1-v) }{m^2-p_\perp^2 v(v-1)}\Pi_\perp^{\mu\nu}.  
\end{equation}
Finally, integrating over $v$, we obtain
\begin{align}
i\Pi^{\mu \nu}(p)&=\frac{e^2|eE|}{2\pi^2}e^{\frac{ip_\parallel^2}{2|eE|}}\nonumber \\
&\times \left(1-\frac{4m^2 \operatorname{arctanh}\left(\frac{p_\perp}{\sqrt{4m^2+p_\perp^2}}\right)}{p_\perp\sqrt{4m^2+p_\perp^2}} \right)\Pi_\perp^{\mu\nu}.  
\label{Pimunulimitinfinity}
\end{align}
We highlight that Eq.(\ref{Pimunulimitinfinity}) exactly matches Eq.(\ref{Pimunustrongfieldfinal}), confirming that our general result exhibits the correct behavior in the strong field limit.

\bibliography{mybibliography}

\end{document}